\documentstyle[preprint,aps,epsf]{revtex}                  
\tightenlines
\begin{document}
\def\non{\nonumber}
\def\be{\begin{eqnarray}}
\def\en{\end{eqnarray}}
\def\la{\langle}
\def\ra{\rangle}
\def\hep{\hat{\varepsilon}}
\def\pr{Phys. Rev.~}
\def\prl{Phys. Rev. Lett.~}
\def\pl{Phys. Lett.~}
\def\np{Nucl. Phys.~}
\def\zp{Z. Phys.~}
\def\bi{\bibitem}
\pagestyle{empty}                                      
\draft
\vfill
\title{Electromagnetic properties of heavy mesons \\
in heavy quark limit}

\author{Chien-Wen Hwang} %
\address{\rm Department of Physics, National Tsing Hua University, Hsinchu 300, Taiwan \rm}
%
%
\vfill
\maketitle
\begin{abstract}
The electromagnetic (EM) form fators of the pseudoscalar and vector mesons are calculated in the light-front framework. We extract these form factors from the relevant matrix elements directly, instead of choosing the Breit frame. The results show that the charge radius of the meson are related to both the first and second longitudinal momentum square derivative of the momentum distribution function. In addition, the static properties of the EM form factors and the heavy quark symmetry of the charge radii are checked analytically when we take the heavy quark limit. 
\end{abstract}
%
%
\pagestyle{plain}
The understanding of the electromagnetic (EM) properties of hadrons is an important topic, and the EM form factors which are calculated within the non-petrturbative methods are the useful tool for this purpose. There have been numerous experimental \cite{Exp0,Exp1,Exp2,Exp3,Exp4,Exp5} and theoretical studies \cite{The0,The1,The2} of the EM form factors of the light pseudoscalar meson ($\pi$ ana $K$). However, duing to the experiments are more difficult, the EM form factors of the light vector meson ($\rho$ and $K^*$) have fewer investigations than their pseudoscalar counterparts \cite {HP,Simula}, even though they could provide much information about the bound-state dynamics. As for the EM form factors of the heavy mesons (which containing one heavy quark), there are much fewer study than the light ones. In the heavy hadron investigation, however, the heavy quark symmetry (HQS) \cite{HQS} is a fundamental and model-independent property. In this work, we will study the EM form factors of the heavy mesons with the light-front framework, and will check whether HQS is satisified or not among these EM properties. 

As is well known \cite{LFQM}, the light-front quark model (LFQM) allows an exact separation in momentum space between the center-of-mass motion and intrisic wave functions, a consistent treatment of quark spins and the center-of-mass motion can also be carried out. Thus it has been applied in the past to calculate various form factors \cite{LFQMa}. Let us begin with the light-front formalism. A meson bound state consisting of a heavy quark $q_1$ and
an antiquark $\bar q_2$ with total momentum $P$
and spin $S$ can be written as
\begin{eqnarray}
        |M(P, S, S_z)\rangle
                &=&\int \{d^3p_1\}\{d^3p_2\} ~2(2\pi)^3 \delta^3(\tilde
                P-\tilde p_1-\tilde p_2)~\nonumber\\
        &&\times \sum_{\lambda_1,\lambda_2}
                \Psi^{SS_z}(\tilde p_1,\tilde p_2,\lambda_1,\lambda_2)~
                |q_1(p_1,\lambda_1) \bar q_2(p_2,\lambda_2)\rangle,
\end{eqnarray}
where $p_1$ and $p_2$ are the on-mass-shell light-front momenta,
\begin{equation}
        \tilde p=(p^+, p_\bot)~, \quad p_\bot = (p^1, p^2)~,
                \quad p^- = {m^2+p_\bot^2\over p^+},
\end{equation}
and
\begin{eqnarray}
        &&\{d^3p\} \equiv {dp^+d^2p_\bot\over 2(2\pi)^3}, \nonumber \\
        &&|q(p_1,\lambda_1)\bar q(p_2,\lambda_2)\rangle
        = b^\dagger_{\lambda_1}(p_1)d^\dagger_{\lambda_2}(p_2)|0\rangle,\\
        &&\{b_{\lambda'}(p'),b_{\lambda}^\dagger(p)\} =
        \{d_{\lambda'}(p'),d_{\lambda}^\dagger(p)\} =
        2(2\pi)^3~\delta^3(\tilde p'-\tilde p)~\delta_{\lambda'\lambda}.
                \nonumber
\end{eqnarray}
In terms of the light-front relative momentum
variables $(x, k_\bot)$ defined by
\begin{eqnarray}
        && p^+_1=(1-x) P^+, \quad p^+_2=x P^+, \nonumber \\
        && p_{1\bot}=(1-x) P_\bot+k_\bot, \quad p_{2\bot}=x P_\bot-k_\bot,
\end{eqnarray}
the momentum-space wave-function $\Psi^{SS_z}$
can be expressed as
\begin{equation}
        \Psi^{SS_z}(\tilde p_1,\tilde p_2,\lambda_1,\lambda_2)
                = R^{SS_z}_{\lambda_1\lambda_2}(x,k_\bot)~ \phi(x, k_\bot), \label{momentumspace}
\end{equation}
where $\phi(x,k_\bot)$ describes the momentum distribution of the
constituents in the bound state, and $R^{SS_z}_{\lambda_1\lambda_2}$
constructs a state of definite spin ($S,S_z$) out of light-front
helicity ($\lambda_1,\lambda_2$) eigenstates.  Explicitly,
\begin{equation}
        R^{SS_z}_{\lambda_1 \lambda_2}(x,k_\bot)
                =\sum_{s_1,s_2} \langle \lambda_1|
                {\cal R}_M^\dagger(1-x,k_\bot, m_1)|s_1\rangle
                \langle \lambda_2|{\cal R}_M^\dagger(x,-k_\bot, m_2)
                |s_2\rangle
                \langle {1\over2}s_1;
                {1\over2}s_2|S,S_z\rangle, \label{RR}
\end{equation}
where $|s_i\rangle$ are the usual Pauli spinors,
and ${\cal R}_M$ is the Melosh transformation operator \cite{Melosh}:
\begin{equation}
        {\cal R}_M (x,k_\bot,m_i) =
                {m_i+x M_0+i\vec \sigma\cdot\vec k_\bot \times \vec n
                \over \sqrt{(m_i+x M_0)^2 + k_\bot^2}}, \label{Melosh2}
\end{equation}
with $\vec n = (0,0,1)$, a unit vector in the $z$-direction, and
\begin{equation}
        M_0^2={ m_1^2+k_\bot^2\over (1-x)}+{ m_2^2+k_\bot^2\over x}.
\label{M0}
\end{equation}
In practice it is more convenient to use the covariant form for
$R^{SS_z}_{\lambda_1\lambda_2}$ \cite{Jaus}:
\begin{equation}
        R^{SS_z}_{\lambda_1\lambda_2}(x,k_\bot)
                ={\sqrt{p_1^+p_2^+}\over \sqrt{2} ~{\widetilde M_0}}
        ~\bar u(p_1,\lambda_1)\Gamma v(p_2,\lambda_2), \label{covariant}
\end{equation}
where
\begin{eqnarray}
        &&{\widetilde M_0} \equiv \sqrt{M_0^2-(m_1-m_2)^2}, \nonumber\\
        &&\Gamma=\gamma_5 \qquad ({\rm pseudoscalar}, S=0). \non \\
        &&\Gamma=-\not{\! \hep}(S_z)+{\hep\cdot(p_1-p_2)
        \over M_0+m_1+m_2} \qquad ({\rm vector}, S=1), \nonumber
\end{eqnarray}
with
\begin{eqnarray}
        &&\hep^\mu(\pm 1) =
                \left[{2\over P^+} \vec \varepsilon_\bot (\pm 1) \cdot
                \vec P_\bot,\,0,\,\vec \varepsilon_\bot (\pm 1)\right],
                \quad \vec \varepsilon_\bot
                (\pm 1)=\mp(1,\pm i)/\sqrt{2}, \nonumber\\
        &&\hep^\mu(0)={1\over M_0}\left({-M_0^2+P_\bot^2\over
                P^+},P^+,P_\bot\right).   \label{polcom}
\end{eqnarray}
We normalize the meson state as
\begin{equation}
        \langle M(P',S',S'_z)|M(P,S,S_z)\rangle = 2(2\pi)^3 P^+
        \delta^3(\tilde P'- \tilde P)\delta_{S'S}\delta_{S'_zS_z}~,
\label{wavenor}
\end{equation}
so that the normalization condition of the momentum distribution function can be obtained
\begin{equation}
        \int {dx\,d^2k_\bot\over 2(2\pi)^3}~|\phi(x,k_\bot)|^2 = 1. 
\label{momnor}
\end{equation}
In principle, $\phi(x,k_\bot)$ can be obtained by solving the light-front QCD bound state equation\cite{Zhang}. However, such first-principles solutions have not been done yet, and the phenomenological wave functions must be considered. In this paper, we don't designate any one to use but only ask the momentum distribution function has the scaling behavior \cite{Cheung}
\be
\phi_{Q \bar{q}}(x,k_\perp)\to~\sqrt{m_Q}\,\Phi(m_Qx,k_\perp),  \label{hqamp}
\en
in the infinite quark mass limit $m_Q \to \infty$. The factor $\sqrt{m_Q}$ comes from the particular normalization we have assumed for the physical state in Eqs. (\ref{wavenor}) and (\ref{momnor}). The reason why the light-front heavy-meson wave function should have such an asymptotic form is as follows. Since $x$ is the longitudinal momentum fraction carried by the light 
antiquark, the meson wave function should be sharply peaked near $x\sim\Lambda_{\rm QCD}/m_Q$. It is thus clear that only terms of the form``$m_Qx$" survive in the wave function as $m_Q\to\infty$; that is, $m_Qx \equiv X$ is independent of $m_Q$ in the $m_Q\to\infty$ limit.

Now we consider the pseudoscalar meson first. The EM form factor of a meson $P$ is determined by the scattering of one virtual photon and one meson. Since the momentum of the virtual photon $q$ is space-like, it is always possible to orient the axes in such a manner that $Q^+=0$. Thus the EM form factor of a pseudoscalar meson $P$ is determined by the matrix element
\be
\la P(P')|J^+|P(P)\ra = e~F_P(Q^2) (P+P')^+, \label{FPdef}
\en
where $J^\mu = \bar q e_q e\gamma^\mu q$ is the vector current, $e_q$ is the charge of quark $q$ in $e$ unit, and $Q^2=(P'-P)^2\leq 0$. This process is illustrated in Fig.1. With LFQM, $F_P$ can be extracted by Eq. (\ref{FPdef})
\be
F_P(Q^2)&=&e_{q_1}\int {dxd^2k_\perp\over{2(2\pi)^3}}{\phi_P(x,k_\perp)\over{\sqrt{{\cal A}^2+k^2_\perp}}}{\phi_{P'}(x,k'_\perp)\over{\sqrt{{\cal A}^2+k'^2_\perp}}}\left[{\cal A}^2+k_\perp\cdot k'_\perp\right]\non \\
&+&e_{\bar {q_2}}\int {dxd^2k_\perp\over{2(2\pi)^3}}{\phi_P(x,k_\perp)\over{\sqrt{{\cal A}^2+k^2_\perp}}}{\phi_{P'}(x,k''_\perp)\over{\sqrt{{\cal A}^2+k''^2_\perp}}}\left[{\cal A}^2+k_\perp\cdot k''_\perp\right], \label{FPgeneral}
\en
where $k'_\perp=k_\perp+x Q_\perp$, $k''_\perp=k_\perp-(1-x) Q_\perp$, ${\cal A}=xm_1+(1-x)m_2$, and $\sqrt{{\cal A}^2+k_\perp^2}=\widetilde{M}_0\sqrt{x(1-x)}$. From Eqs. (\ref{RR}), (\ref{Melosh2}), and (\ref{covariant}), it is understandable that the term $\sqrt{{\cal A}^2+k_\perp^2}$ comes from the Melosh transformation. After fixing the parameters which appear in the wave fnction, Eq. (\ref{FPgeneral}) fits the experimental data well \cite{hwang2}. But this is not the whole story. We consider the term $\widetilde {\phi}_{P} \equiv {\phi_{P}(x,k_\perp)/{\sqrt{{\cal A}^2+k^2_\perp}}}$ and take the Tayor expansion around $k^2_\perp$
\be
\widetilde {\phi}_{P'}(k'^2_\perp)=\widetilde {\phi}_{P'}(k^2_\perp)+{d\widetilde {\phi}_{P'}\over{dk^2_\perp}}\Bigg|_{Q_\perp=0}(k'^2_\perp-k^2_\perp)+{d^2\widetilde {\phi}_{P'}\over{2(dk^2_\perp)^2}}\Bigg|_{Q_\perp=0}(k'^2_\perp-k^2_\perp)^2+.....
\en
Then, by using the idenity
\be
\int d^2k_\perp~(k_\perp \cdot Q_\perp)(k_\perp \cdot Q_\perp)={1\over{2}}\int d^2k_\perp~k^2_\perp~Q^2_\perp, \label{QQQ2}
\en
we can rewrite (\ref{FPgeneral}) to
\be
F_P(Q^2)&=&(e_{q_1}+e_{\bar q_2})\non \\
&-&Q^2 \int{dxd^2k_\perp\over{2(2\pi)^3}}\phi^2_P(x,k_\perp)(e_{q_1}x^2+e{\bar q_2}(1-x)^2)\Bigg( \Theta_{P}{{\cal A}^2+2k^2_\perp\over{{\cal A}^2+k^2_\perp}}+\widetilde{\Theta}_{P}k^2_\perp\Bigg)\non \\
&+&{\cal O}(Q^4), \label{FPQQ}
\en
where
\be
\Theta_{M}={1\over{\widetilde {\phi}_{M}}}\Bigg({d\widetilde {\phi}_{M}\over{dk^2_\perp}}\Bigg),~~\widetilde{\Theta}_{P}={1\over{\widetilde {\phi}_{M}}}\Bigg({d^2\widetilde {\phi}_{M}\over{(dk^2_\perp)^2}}\Bigg).
\en
The mean square radius $\la r^2 \ra_P\equiv 6(dF_P(Q^2)/dQ^2)|_{Q^2=0}$ can be obtained easily that $\la r^2 \ra_P \equiv \la r^2 \ra_{q_1}+\la r^2 \ra_{{\bar q}_2}$, where
\be
\la r^2 \ra_{q_1}&=&e_{q_1}\Bigg\{-6\int{dxd^2k_\perp\over{2(2\pi)^3}}x^2\widetilde {\phi}_{P}\Bigg[({\cal A}^2+2k^2_\perp){d\over{dk^2_\perp}}+({\cal A}^2+k^2_\perp)k^2_\perp\Bigg({d\over{dk^2_\perp}}\Bigg)^2\Bigg]\widetilde {\phi}_{P}\Bigg\},\label{MSRq1} \\
\la r^2 \ra_{\bar{q}_2}&=&e_{\bar{q}_2}\Bigg\{-6\int{dxd^2k_\perp\over{2(2\pi)^3}}(1-x)^2\widetilde {\phi}_{P}\Bigg[({\cal A}^2+2k^2_\perp){d\over{dk^2_\perp}}+({\cal A}^2+k^2_\perp)k^2_\perp\Bigg({d\over{dk^2_\perp}}\Bigg)^2\Bigg]\widetilde {\phi}_{P}\Bigg\}.\label{MSRq2}
\en
It is worthwhile to mention that, first, the static property $F_P(0)=e_P$ is quite easily checked in (\ref{FPQQ}). Secondly, from Eqs. (\ref{MSRq1}) and (\ref{MSRq2}), we find that the mean square radius is related to the first and second longitudinal momentum square derivatives of $\widetilde {\phi}$ which contain the Melosh transformation effect. When we consider the heavy meson $q_1 \to Q$, we get
\be
x \to {X\over{m_Q}},~~{\cal A} \to \widetilde {\cal A} \equiv X+m_2, \label{hqlxa}
\en
and Eq.(\ref{hqamp}) in the heavy quark limit ($m_Q \to \infty$), then the mean square radius
\be
\la r^2 \ra_Q &=&e_Q\Bigg\{{-6\over{m^2_Q}} \int{dXd^2k_\perp\over{2(2\pi)^3}}X^2\widetilde {\Phi}\Bigg[(\widetilde {\cal A}^2+2k^2_\perp){d\over{dk^2_\perp}}+(\widetilde {\cal A}^2+k^2_\perp)k^2_\perp\Bigg({d\over{dk^2_\perp}}\Bigg)^2\Bigg]\widetilde {\Phi}\Bigg\} \non \\
&\to& 0, \label{HQL1} \\
\la r^2 \ra_{\bar {q}_2} &=&e_{\bar {q}_2}\Bigg\{-6 \int{dXd^2k_\perp\over{2(2\pi)^3}}\widetilde {\Phi}\Bigg[(\widetilde {\cal A}^2+2k^2_\perp){d\over{dk^2_\perp}}+(\widetilde {\cal A}^2+k^2_\perp)k^2_\perp\Bigg({d\over{dk^2_\perp}}\Bigg)^2\Bigg]\widetilde {\Phi}\Bigg\},\label{HQL2}
\en
where $\widetilde {\Phi} ={\Phi}/\sqrt{\widetilde {\cal A}^2+k^2_\perp}$. Eq. (\ref{HQL1}) means that, in the heavy quark limit, the mean square radius $\la r^2 \ra_P$ is blind to the flavor of $Q$. This is the so-called flavor symmetry.

Next, the vector meson case is considered. As in the pseudoscalar case, we set $Q^+=0$, then the EM form factor of a vector meson $V$ is parametrized by the matrix element
\be
\la V'(P',\epsilon')|J^+|V(P,\epsilon)\ra&=&G_1(Q^2)\epsilon\cdot \epsilon'(P+P')^+\non \\
&-&G_2(Q^2)(P\cdot \epsilon'~\epsilon^++P'\cdot \epsilon~\epsilon'^+)\non \\
&+&G_3(Q^2){1\over{2M_V^2}}P\cdot \epsilon'~P'\cdot \epsilon (P+P')^+,
\en
where $M_V$ is the V-meson mass and $\epsilon (\epsilon')$ is the polorization of $V (V')$. In the literature \cite{CCKP,FFS,GK,BH}, these invariant form factors are usually obtained by choosing the Breit frame. For the vector meson there are only three independent form factors, whereas four matrix elements of $J^+$ are independent \cite{The0} since $J^+$ is (i) Hermitian, (ii) invariant under rotations about $\vec n$, (iii) time reveral, and (iv) reflection on the plane perpendicular to $\vec n$. An additional constraint, which called the angular condition, is related to the rotational invariance of the charge density \cite{Simula}. This invariance is related to the unitary transformations based upon a subset of Poincare generators. The authors of Ref. \cite{GK} had written the angular condition in the following form
\be
\Delta (Q^2) \equiv (1+2 \eta)J_{1,1} +J_{1,-1}-\sqrt{8\eta}J_{1,0}-J_{0,0}=0,\label {angular}
\en
where $\eta=Q^2/4M^2_V$ and $J_{\epsilon',\epsilon} \equiv \la V'(\vec {P'},\epsilon')| J^+|V(\vec {P},\epsilon)\ra$. In generally, this condition must be satisfied when the matrix elements of the many-body current are existence, and are breakdown for the one-body one alone. Thus, when only the one-body current operator $\bar q \gamma^\mu q$ is considered, Eq.(\ref{angular}) usually doesn't equal to zero and the calculation of these form factors depends upon the description used \cite{Simula}. Following \cite{BH,LB}, $J_{0,0}$ is expected to be the dominant matrix element in the pQCD regime. Thus, as the arguement in \cite{GK}, it is reasonable that we calculate the $G_i$ form factor without using the matrix element $J_{0,0}$ in the constituent quark model.

In this paper, we will extract the form factors directly, instead of choosing the Breit frame. The $Q^2$ dependence of these form factors must be calculated order by order. In spite of the fact that we cannot obtained the exact form, the advantages, however, are fruitful and will be mentioned later. We choose the matrix elements $J_{1,1}$, $J_{1,-1}$, $J_{1,0}$ and Eq. (\ref{QQQ2}) to extract the form factors $G_1$, $G_2$, and $G_3$ in the suitable order
\be
G_1(Q^2)&=&(e_{q_1}+e_{\bar {q}_2})\non \\
&-&Q^2 \int{dxd^2k_\perp\over{2(2\pi)^3}}\widetilde {\phi}^2_{V}\Bigg\{e_{q_1}\Bigg[x^2\Theta_V({\cal A}^2+2 k^2_\perp)+x^2\widetilde{\Theta}_V k^2_\perp({\cal A}^2+k^2_\perp)\non \\
&&~~~~~~~~~~~~~~~~~~~~~~~~~~~~-{k^2_\perp\over{M^2_V}}+{x k^4_\perp\over{2 (1-x)M^2_0 W^2_V}}\Bigg]+e_{\bar {q}_2}\Bigg[x \leftrightarrow (1-x)\Bigg]\Bigg\} \non \\
&+&O(Q^4), \\
{G_2(Q^2)+2 \over{M_V}}&=&\int{dxd^2k_\perp\over{2(2\pi)^3}}\widetilde {\phi}^2_{V}\Bigg\{e_{q_1}\Bigg[x a+x^2 \Bigg(b+{d\over{W_V}}\Bigg)-{x k^2_\perp\over{2 W_V}}\Bigg(c+{d\over{(1-x)M_0 W_V}}\Bigg)\non \\
&&~~~~~~~~~~~~~~~~~~~~~~~~+x k^2_\perp \Theta_V \Bigg(b+{d\over{W_V}}\Bigg)\Bigg]+e_{\bar {q}_2}\Bigg[x \leftrightarrow (1-x)\Bigg]\Bigg\}\non \\
&+&O(Q^2), \\
{G_3(Q^2) \over{4M^2_V}}&=&\int{dxd^2k_\perp\over{2(2\pi)^3}}(e_{q_1}x^2+e_{\bar {q}_2}(1-x)^2)\widetilde {\phi}^2_{V}\Bigg\{-{{\cal A}\over {W_V}}+{{\cal A}k^2_\perp\over{2 x (1-x)M_0 W^2_V}}\Bigg\}\non \\
&+&O(Q^2),
\en
where $W_V=M_0+m_1+m_2$ and
\be
a&=&{1\over{M_0}}\Bigg[{x\over{1-x}}(m^2_1+k^2_\perp)-{1-x\over{x}}(m^2_2+k^2_\perp)+\widetilde {M}^2_{0}(1-2 x)-{2{\cal A}\over{W_V}}(M^2_0-(m^2_1+m^2_2))\Bigg],\non \\
b&=&{-4\over{M_0}}\Bigg[{x\over{1-x}}(m^2_1+k^2_\perp)-{1-x\over{x}}(m^2_2+k^2_\perp)-{m_1+m_2\over{W_V}}(M^2_0-(m^2_1+m^2_2))\Bigg],\non \\
c&=&4\Bigg[{m_1-m_2\over{M_0}}-{M_0\over{W_V}}+{m^2_1+m^2_2\over{M_0 W_V}}\Bigg],\non \\
d&=&4{1-x\over{M_0}}\Bigg[(m_1+m_2)\Bigg({x\over{1-x}}(m^2_1+k^2_\perp)-{1-x\over{x}}(m^2_2+k^2_\perp)\Bigg)\non \\
&&~~~~~~~~~+(M_0-m_1-m_2)(M^2_0-(m^2_1+m^2_2))\Bigg].
\en
These $G_i (Q^2)$ form factors are related to the charge, magnetic and quadrupole form factors as \cite{HP}
\be
G_C(Q^2)&=&G_1+{2\over{3}}{Q^2\over{4M_V^2}}G_Q, \non \\
G_M(Q^2)&=&-G_2, \non \\
G_Q(Q^2)&=&G_1+G_2+\Bigg(1+{Q^2\over{4M^2_V}}\Bigg)G_3.
\en
These form factors are proportional to the static quantities of charge $e$, magnetic moment $\mu$, and quadrupole moment $Q$ when $Q^2=0$,
\be
G_C(0)&=& e_V, \non \\
G_M(0)&=& \mu,~(\text{the unit is}~e/(2M_V)), \non \\
G_Q(0)&=& Q~M_V^2.
\en
The mean square radius of the vector meson $\la r^2 \ra_V=6(dG_C(Q^2)/dQ^2)|_{Q^2=0}$ is
\be
\la r^2 \ra_V&=&\la r^2 \ra_{q_1}+\la r^2 \ra_{\bar {q}_2} \non \\
&=&e_{q_1}\Bigg\{-6\int{dxd^2k_\perp\over{2(2\pi)^3}}x^2\widetilde {\phi}_{V}\Bigg[({\cal A}^2+2k^2_\perp){d\over{dk^2_\perp}}+({\cal A}^2+k^2_\perp)k^2_\perp\Bigg({d\over{dk^2_\perp}}\Bigg)^2\non \\
&&~~~~~~~~~~~~~~~~~~~~~~~~~~~~~~~~-{k^2_\perp\over{(xW_V)^2}}+{k^4_\perp\over{2x(1-x)M^2_0 W^2_V}}\Bigg]\widetilde {\phi}_{V}\Bigg\} \non  \\
&+&e_{\bar{q}_2}\Bigg\{-6\int{dxd^2k_\perp\over{2(2\pi)^3}}(1-x)^2\widetilde {\phi}_{V}\Bigg[({\cal A}^2+2k^2_\perp){d\over{dk^2_\perp}}+({\cal A}^2+k^2_\perp)k^2_\perp\Bigg({d\over{dk^2_\perp}}\Bigg)^2\non \\
&&~~~~~~~~~~~~~~~~~~~~~~~~~~~~~~~~~~~~~~~-{k^2_\perp\over{((1-x)W_V)^2}}+{k^4_\perp\over{2x(1-x)M^2_0 W^2_V}}\Bigg]\widetilde {\phi}_{V}\Bigg\}\non \\
&+&{1\over{M^2_V}}(G_1+G_2+G_3)\Bigg|_{Q^2=0}\label{MSRV}
\en
Similar to the pseudoscalar case, in addition to Eqs. (\ref{hqamp}) and (\ref{hqlxa}), there are some behaviors existing
\be
M_0 \to m_Q + O(m_Q x),~~~~~W_V \to 2m_Q+ O(m_Q x),
\en
if the heavy quark limit is taken. We also obtain the Eqs. (\ref{HQL1}) and (\ref{HQL2}) for vector meson case. Thus the mean square radii of pseudoscalar and vector meson are equal $\la r^2 \ra_P =\la r^2 \ra_V$. This means that they satisfy the spin symmetry. Combine this relation and Eq.(\ref{HQL1}), we find that the light degrees of freedom are blind to the flavor and spin orientation of the heavy quark. This is the so-called heavy quark symmetry (HQS). Up to now, we have not designated any wave function yet, this also satisfies the well-known property that HQS is model-independent. Reviewing the processes, we can realize that, in this approach, the static properties of the EM form factors and the heavy quark symmetry of the mean square radii can be checked much more easily than in the Breit frame. This is the major reason why we calculate the $Q^2$ dependence of those form factors order by order.

In conclusion, we have calculated the EM form factors of the pseudoscalar and vector mesons. Specially, the EM form factors of vector meson are extracted from the relevant matrix elements directly, instead of choosing the Breit frame. We found that the charge radius is related to both the first and second longitudinal momentum square derivative of the momentum distribution function. We also found the static properties of the EM form factors and the heavy quark symmetry of the mean square radii are checked analytically by evaluating the $Q^2$ dependence of those form factors order by order. Therefore, in the heavy quark limit, the charge radii of pseudoscalar and vector mesons have both spin and flavor symmetries, and these properties are model-independent. 
\acknowledgments
This work was supported in part by the National Science Council of 
R.O.C. under the Grant No. NSC89-2112-M-007-054.


\newpage
\parindent=0 cm
\centerline{\bf FIGURE CAPTIONS}
\vskip 0.5 true cm

{\bf Fig. 1 } Diagrams contributing to the EM form factors of meson.
\vskip 0.25 true cm

\newpage

\begin{figure}[h]
\hskip 5.7cm
\hbox{\epsfxsize=7.6cm
      \epsfysize=15cm
      \epsffile{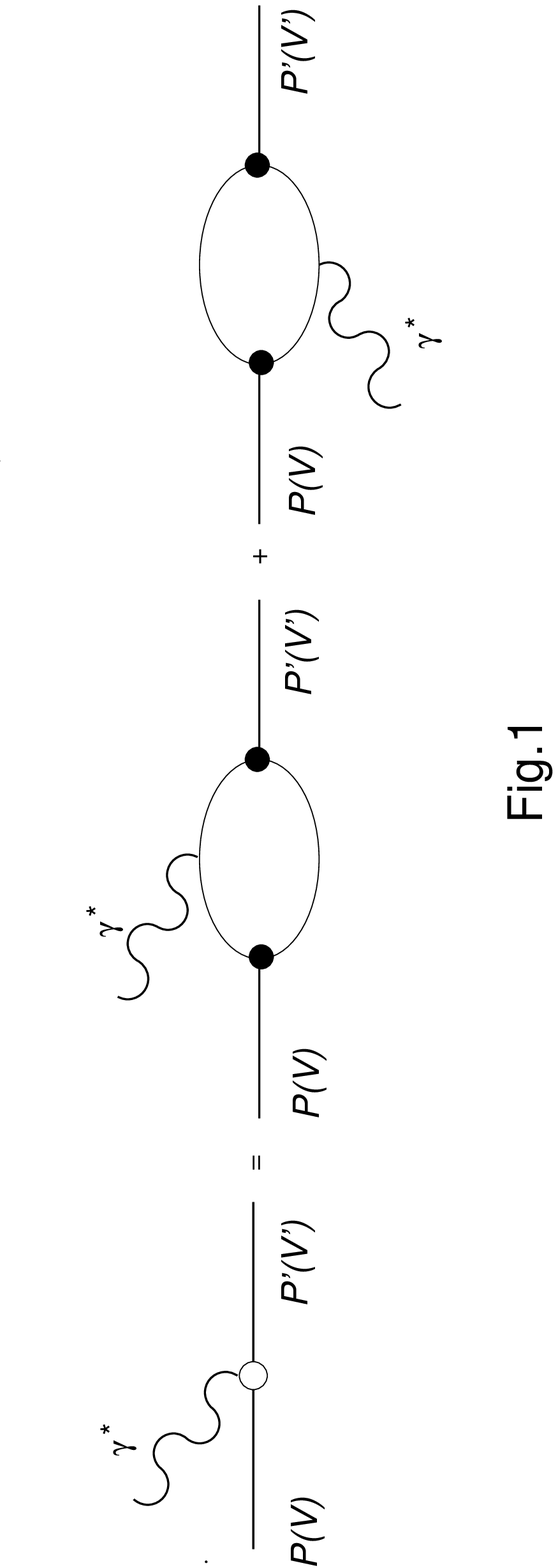}}
\end{figure}
\end{document}